\documentclass[aps,twocolumn,prb,reprint,superscriptaddress,noshowpacs,nofootinbib,noshowkeys,floatfix,longbibliography]{revtex4-2}
\usepackage[colorlinks=true,linktocpage=true,linkcolor=blue,citecolor=blue]{hyperref}
\usepackage[usenames,dvipsnames]{color}
\usepackage{amsmath, amssymb}
\usepackage{multirow}
\usepackage{longtable}
\usepackage{comment}
\usepackage{color}
\usepackage{xcolor}
\usepackage{xspace}
\usepackage{cleveref}
\usepackage[normalem]{ulem}  % \sout{old text} for strikeout
\usepackage{graphicx,lipsum}
\usepackage{ragged2e}
\setcitestyle{numbers}
\setcitestyle{square}

\begin{document}

\title{Evidence of magnetoelastic coupling and magnetic phase coexistence in Mn$_{1.7}$Fe$_{1.3}$Si Heusler Alloy}

\author{Kulbhushan Mishra}
\affiliation{Indian Institute of Technology, Khandwa Road, Indore, Simrol, 453552, India}
\author{Elaine T. Dias}
\affiliation{School of Physical and Applied Sciences, Goa University, Taleigao Plateau, Goa, India, 403206}
\author{Rajeev Joshi}
\affiliation{UGC-DAE Consortium for Scientific Research, University Campus, Khandwa Road, Indore 452001, India}
\author{A. D. Fortes}
\affiliation{ISIS Pulsed Neutron and Muon Source, STFC Rutherford Appleton Laboratory, Harwell Campus, Didcot, Oxon OX11 0QX, United Kingdom}
\author{Christopher M. Howard}
\affiliation{ISIS Pulsed Neutron and Muon Source, STFC Rutherford Appleton Laboratory, Harwell Campus, Didcot, Oxon OX11 0QX, United Kingdom}
\author{Rajeev Rawat}
\affiliation{UGC-DAE Consortium for Scientific Research, University Campus, Khandwa Road, Indore 452001, India}
\author{P. A. Bhobe}
\email[Corresponding author: ]{pbhobe@iiti.ac.in}
 \affiliation{Indian Institute of Technology, Khandwa Road, Indore, Simrol, 453552, India}

\begin{abstract}
Noncollinear metallic antiferromagnets, with their rapid spin dynamics, efficient spin transport, and distinctive spin textures, play a pivotal role in advancing the field of spintronics. In this study, we report a comprehensive investigation of the structural, magnetic, and transport properties of cubic Mn$_{1.7}$Fe$_{1.3}$Si Heusler compound. Temperature-dependent magnetization measurement reveals a paramagnetic to ferromagnetic transition at $T_C$ = 85 K, followed by a spin reorientation transition. Neutron diffraction data, analyzed as a function of temperature, demonstrates that the occurrence of a spin-reorientation transition is accompanied by magnetoelastic coupling, as evidenced by a change in unit cell volume below $T_C$. Magnetic structure refinement of the low-temperature neutron powder diffraction data confirms the canted antiferromagnetic ordering below 55 K. The metallic nature of the sample is confirmed by the gradual decrease in the $\rho$(T) with decreasing temperature. At low temperatures, a field-induced metamgnetic transition is observed in both, magnetization and magneto-transport measurements. The $H-T$ phase diagram shows a phase-coexistence region emerging at low temperatures for H $<$ 2.5 T. These findings provide valuable insights into the magnetic and transport behavior of the Heusler compounds, underscoring their potential for spintronic applications.

\end{abstract}

\maketitle
%\linenumbers
\section{Introduction}
Antiferromagnets have recently drawn much attention from spintronic researchers because of their fascinating magnetotransport properties, robustness against external perturbation, and fast dynamics\cite{AFMreview}. Of particular interest are antiferromagnets with noncoplanar, noncollinear spin structures, as they exhibit a pronounced topological Hall effect (THE) and are known to host magnetic skyrmions.\cite{AFM_TMR, VNbS2_TAH, Mn3Sn_1, Mn3Sn_2}. Synthetic antiferromagnets, where two ferromagnets coupled antiparallel, usually by  Ruderman-Kittel-Kasuya-Yoshida interactions, are currently used to overcome device malfunction associated with the ferromagnetic stray field when lateral dimensions are reduced. The wider class of insulating antiferromagnets, which include MnO, NiO, and CoO, found their application in spin Seebeck\cite{Seki-2015}, tunneling magnetoresistance\cite{TMR}, multiferroicity, and magnetoelectric effects\cite{Martin_2008}. However, due to the absence of free charge carriers, spin transport in insulating antiferromagnets is limited to magnonic excitations (quanta of spin waves), which are less efficient for spintronic applications. 

In this context, metallic antiferromagnets, particularly Mn-based alloys like FeMn, IrMn, Mn$_2$Au, and PtMn, have become the focal point of research. These materials are extensively utilized in applications such as read heads for hard disk drives and magnetic memory devices.\cite{parkin1991}. The noncollinear spin texture of $\gamma$--FeMn and Mn$_3$Ir breaks the invariance under the combination of time-reversal symmetry with crystal symmetry, resulting in a finite anomalous Hall effect\cite{Nagaosa-2001, MacDonald-2014}. The inverse galvanic effects have been predicted and experimentally demonstrated in Mn$_2$Au\cite{wangler-2017}. Metallic antiferromagnets also comprise archetypal materials such as Cr which are thoroughly studied in Cr/MgO-based multilayers. Additionally, metallic metamagnets with an antiferromagnetic to ferromagnetic/ferrimagnetic transition, such as FeRh, offer the possibility to indirectly operate the antiferromagnetic order via iterative steps consisting of manipulating the ferromagnetic order and undergoing the magnetic phase transition\cite{FeRh-1, FeRh-2}. Other important materials include, for example, Gd alloys, such as GdSi, GdGe, and GdAu$_2$ \cite{Gd-1996}.

Among intermetallic compounds, the Heusler family stands out for its intriguing characteristics, which vary depending on the doping and disorder. These include half-metallicity\cite{HM1, HM2}, shape memory effect\cite{SME}, the magnetocaloric effect\cite{PAB_apl}, and anomalous hall effect. The transport properties are also significantly influenced by doping-induced changes in electronic and magnetic structures. Within the Heusler family, Fe$_{3-x}$Mn$_x$Si\cite{FMS_1} is a notable system, which shows a long-range ferromagnetic order above ambient temperature for $x$ $\leq$ 0.75. However, as Mn concentration decreases ($x$ $\geq$ 0.75), the ferromagnetic ordering temperature decreases due to reduced exchange interaction between Fe atoms at different sites. Theoretically, this system exhibits half-metallicity for $x$ $\geq$ 0.75\cite{FMS_SP}. At low temperatures, field-induced irreversibility in magnetoresistance(MR) is observed for higher Mn concentration ($x$ $\geq$ 0.9)\cite{Lpal}. Masahiko \textit{et al.} studied Fe$_{3-x}$Mn$_x$Si for 0.75 $\leq$ $x$ $\leq$ 1.9 and found that there is an increase in antiferromagnetic (AFM) transition temperature while the paramagnetic to ferromagnetic transition temperature systematically decreases\cite{Mn1.7_1}. A few specific compositions of this series, particularly $x$ = 1.6, 1.7, 1.8, are of immense interest\cite{Mn1.7_2}. Specifically, $x$ = 1.7, there are two AFM transitions around 65 K ($T_{N1}$) and 55 K ($T_{N2}$), in addition to a ferromagnetic transition around 85 K. The second transition temperature, $T_{N2}$, decreases with increasing applied magnetic field. Additionally, transport measurement reveals a change in the slope of the field-dependent magnetoresistance from positive to negative at low temperatures\cite{Mn1.7_3}.  

The complex magnetic behavior of Mn$_{1.7}$Fe$_{1.3}$Si and its influence on transport properties remain poorly understood. To date, no detailed study has explored the relation between crystal structure and magnetic properties to explain the observed peculiarities in the physical properties of the Mn$_{1.7}$Fe$_{1.3}$Si. This study comprehensively investigates the magnetotransport, magnetic, and structural properties of Mn$_{1.7}$Fe$_{1.3}$Si. Temperature-dependent resistivity measurement confirms the metallic nature of the prepared composition. The isothermal field-dependent magnetization and magnetotransport data reveal a field-induced metamagnetic transition at low temperatures. Additionally, temperature-dependent magnetization shows the spin reorientation transition (SRT) between 65 K and 55 K. Below 55 K, temperature-dependent neutron powder diffraction (NPD) analysis confirms the presence of canted antiferromagnetic ordering. The variation of unit cell volume with temperature, extracted from the refinement of the NPD data, shows anomalous slope change around magnetic transitions, hinting at strong magnetoelastic coupling in Mn$_{1.7}$Fe$_{1.3}$Si. The $H-T$ phase diagram, constructed from magnetoresistance and magnetization curves, reveals the presence of magnetic phase coexistence at low temperatures.

\section{Experimental method}
A polycrystalline bead of Mn$_{1.7}$Fe$_{1.3}$Si was prepared using an arc melting furnace. The constituent elements of purity $\geq$ 99.99$\%$ were placed in a water-cooled copper hearth and melted several times in a high-purity Argon atmosphere. The arc-melted sample was heat treated at 800$^{\circ}$C for 24 h in an evacuated quartz tube (pressure $\leq$ 10$^{-5}$ mbar) and then cooled to 600$^{\circ}$C at a rate of 40$^{\circ}$C/hour, before being quenched in water. An additional 2$\%$ Mn was added to the final composition before melting to compensate for the losses during the melting process. The homogeneity of the sample was checked by field emission scanning electron microscope-energy dispersive spectroscopy (EDS) using Model FE-SEM, JEOL, JSM-7610F plus. X-ray diffraction is performed using Empyrean, Malvern Panalytical X-ray diffractometer with Cu K$\alpha$ radiation($\lambda$ = 1.5404 \AA). Magnetization as a function of temperature and magnetic field is measured using a Quantum Design SQUID magnetometer. A standard four-probe method with a homemade resistivity setup equipped with a superconducting magnet system was used for the electrical resistivity and magnetoresistance measurement. The current through the samples was supplied using the Keithley 6221 current source meter, and voltage was measured using the Keithley 2182 A nanovoltmeter. The temperature-dependent neutron powder diffraction data was measured at the HRPD beamline, Rutherford Appleton Laboratory, UK, using the time-of-flight method. The neutron diffraction patterns were collected in the warming cycle in the temperature range between 20 K to 300 K, covering all the magnetic transition temperatures. Nuclear refinement is performed using FULLPROF suite\cite{FullProf}. The magnetic symmetry analysis of magnetic structure refinement was done using the BasIreps tool implemented within the FULLPROF.  

\section{Results}

\begin{figure}[]
 \includegraphics[width=\linewidth]{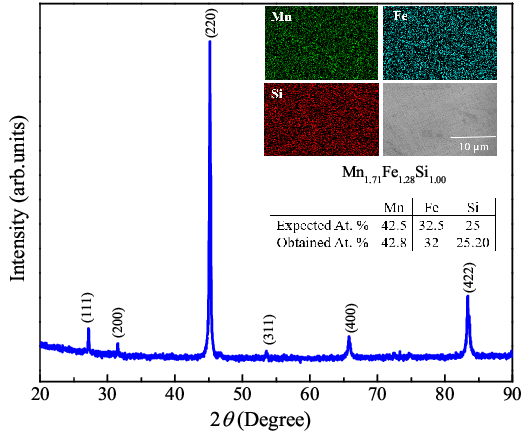}
 \caption{Room temperature X-ray diffraction profile of Mn$_{1.7}$Fe$_{1.3}$Si. The inset of the figure shows the elemental mapping and SEM image at the scale of 10 $\mu$m. Additionally, the table in the inset compares the expected and experimentally obtained atomic percentages from EDX analysis.}
    \label{fig: Fig.1}
    \end{figure}

\begin{figure}[]
 \includegraphics[width=\linewidth]{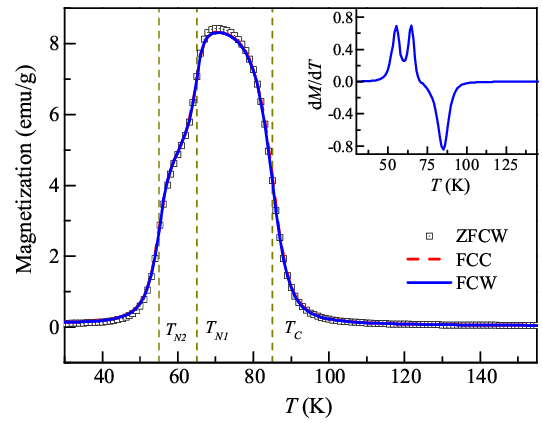}
 \caption{Temperature dependent magnetization, $ M(T)$, of Mn$_{1.7}$Fe$_{1.3}$Si measured at an applied field of 0.01 T.}
    \label{fig: Fig.2}
    \end{figure}

We begin by examining the crystal structure and magnetic transitions of the prepared Mn$_{1.7}$Fe$_{1.3}$Si using X-ray diffraction (XRD), EDS, and temperature-dependent magnetization measurements of the prepared composition. Fig.\ref{fig: Fig.1} shows the room temperature XRD profile of the Mn$_{1.7}$Fe$_{1.3}$Si, where strong superlattice reflections (111) and (200) at low angles indicate the formation of a cubic Heusler structure. No secondary phase is observed in the XRD profile. The inset of Fig.\ref{fig: Fig.1} shows the elemental mapping and SEM image of the Mn$_{1.7}$Fe$_{1.3}$Si, with EDS analysis confirming the Mn, Fe, and Si ratios aligning with the derived stoichiometry. The temperature-dependent magnetization, $M(T)$, measured in an applied external field of 0.01 T, is shown in Fig.\ref{fig: Fig.2}. Measurements were performed under three different protocols: zero-field-cooled warming (ZFCW), field-cooled cooling (FCC), and field-cooled warming (FCW). The $M(T)$ curve undergoes a paramagnetic (PM) to ferromagnetic (FM)/ferrimagnetic (FiM) transition at $T_C$ = 85 K. With a further decrease in temperature, the FM/FiM-to-antiferromagnetic (AFM) transition is marked by a sharp drop in $M$ below 70 K, with two cusp like features at $T_{N1}$ = 65 K and $T_{N2}$ = 55 K. These features are clearly observed in the $dM/dT$ curve, presented in the inset of the figure. All observed transitions match well with the previously reported values\cite{Mn1.7_2, Mn1.7_C}. 

The crystal structure of Heulser alloys is best described in terms of four interpenetrating FCC sublattices situated at four Wyckoff positions (0,0,0), (0.25, 0.25, 0.25), (0.5, 0.5, 0.5) and (0.75, 0.75, 0.75). An ordered Heusler structure, described by compositional formula $X_2YZ$, where $X$ and $Y$ are transition metals and $Z$ is the main group element, crystallizes either with Cu$_2$MnAl-type structure (space group Fm$\bar{3}$m; L2$_1$) or with Hg$_2$CuTi-type structure (space group F$\bar{4}$3m; $XA$). In L2$_1$ structure, the Wyckoff positions (0.75, 0.75, 0.75) and (0.25, 0.25, 0.25) are equivalent and occupied by $X$ atoms, while the positions (0, 0, 0) and (0.5, 0.5, 0.5) are occupied by $Y$ and $Z$ atoms, respectively. On the other hand, in the $XA$ structure, $X$ atom occupies the position (0, 0, 0) and (0.75, 0.75, 0.75), and $Y$ and $Z$ atoms occupy positions (0.25, 0.25, 0.25) and (0.5, 0.5, 0.5), respectively.

\begin{figure}[]
\centering
\includegraphics[width=\linewidth]{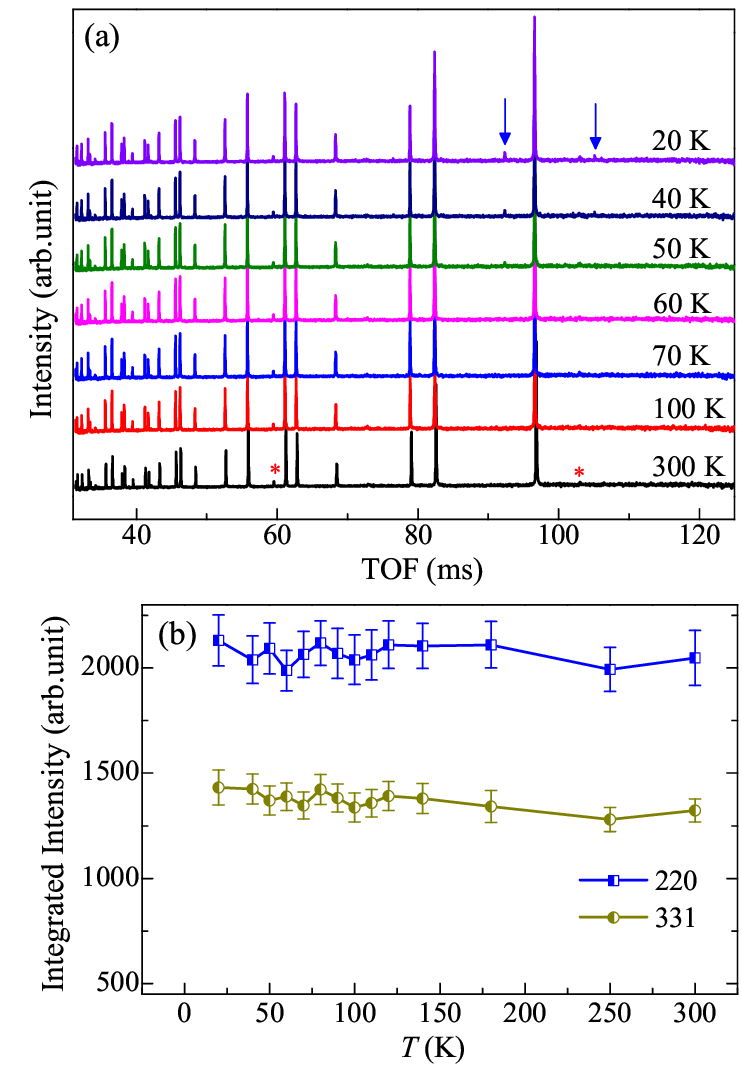}
 \caption{ (a) Temperature-dependent neutron powder diffraction pattern of Mn$_{1.7}$Fe$_{1.3}$Si . The asterisks show spurious peaks from the vanadium window, and arrows mark the new satellite peak below 55 K due to antiferromagnetic ordering. (b) The integrated intensity over the entire temperature range of the two most intense peaks, (220) and (331), present at TOF $\sim$ 96 ms and $\sim$ 82 ms, respectively.}
    \label{fig: Fig.3}
\end{figure}

Amongst the full Heulser compositions derived from the Fe$_{3-x}$Mn$_x$Si series, the Fe$_2$MnSi composition crystallizes in an ordered L2$_1$ structure\cite{Fe2MnSi}, while Mn$_2$FeSi adopts $XA$ structure\cite{Mn2FeSi}. The similar scattering factor of Mn and Fe makes it difficult to distinguish the crystal structure and detect the antisite disorder through XRD analysis. Consequently, the crystal structure of the present composition could be either L2$_1$ or $XA$, a distinction not considered in the previous reports. Furthermore, the two cusp-like features, labeled as $T_{N1}$ and $T_{N2}$ in Fig.\ref{fig: Fig.2}, are often interpreted as the onset of two consecutive antiferromagnetic transitions \cite{Mn1.7_2, Mn1.7_C}. However, to our knowledge, no study has yet fully addressed these magnetic transitions. Previous neutron diffraction measurement on Fe$_{3-x}$Mn$_x$Si was performed above $T_{N1}$, and below $T_{N2}$, missing any potential AFM signal between $T_{N1}$ and $T_{N2}$\cite{Yoon_1977}. To fill this gap in understanding the crystal structure and magnetic ordering in Mn$_{1.7}$Fe$_{1.3}$Si, we performed high-resolution neutron powder diffraction (NPD) measurements. The measurements were conducted from 20 K to 300 K using the time-of-flight (TOF) method, covering all magnetic transitions. Fig.\ref{fig: Fig.3} presents typical data collected from this experiment. At 300 K, the NPD pattern reveals peaks consistent with the cubic full Heusler structure. The peaks marked as asterisks are from vanadium windows in the sample environment. 

\begin{figure}[]
\centering
\includegraphics[width=\linewidth]{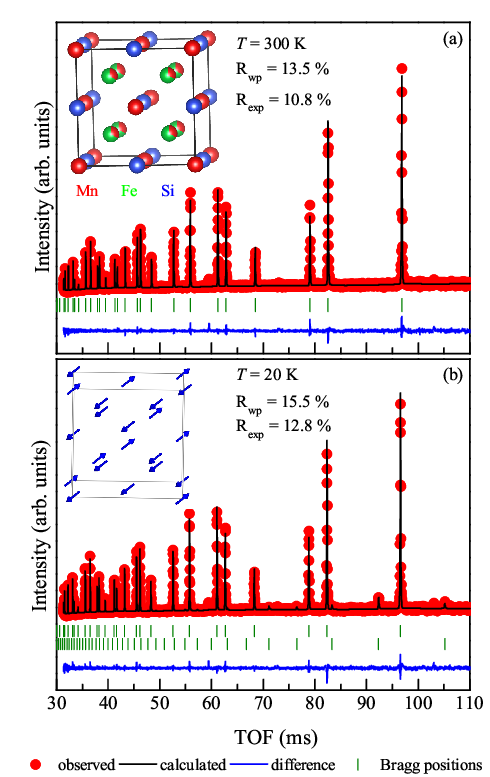}
 \caption{The Rietveld refined neutron powder diffraction pattern of Mn$_{1.7}$Fe$_{1.3}$Si at (a) 300 K and (b) 20 K. The inset in (a) depicts the crystallographic unit cell of Mn$_{1.7}$Fe$_{1.3}$Si, while the inset in (b) illustrates a three-dimensional representation of the magnetic spin arrangement, as determined from the neutron diffraction data at 20 K.}
    \label{fig: Fig.4}
\end{figure}

For analysis of the NPD data, we initially assumed a structural model based on pure XA phase, and space group F$\bar{4}$3m was assumed for the Mn$_{1.7}$Fe$_{1.3}$Si compound at the start of the refinement. In this arrangement, the Wyckoff position (0.25, 0.25, 0.25) is partially substituted by Mn (denoted as Mn1) and Fe (denoted as Fe1). Meanwhile, the remaining Mn (Mn2) and Fe (Fe2) occupy the Wyckoff position (0, 0, 0) and (0.75, 0,75, 0.75). The Si atom occupies the position (0.5, 0.5, 0.5). To account for any antisite disorder, the occupancies were adjusted to preserve the consistent total atom occupancy at both sites. After refining this model, the general formula of the prepared composition turns out to be Mn$_{1.4}$Fe$_{1.68}$Si, implying a straight deviation from the target stoichiometry of Mn$_{1.7}$Fe$_{1.3}$Si. Also, such an atomic ratio contradicts the EDX results, suggesting the prepared composition does not crystallize in the $XA$ phase. As a result, we have modified our model to the L2$_1$ structure (space group Fm$\bar{3}$m) to refine the nuclear structure. In this model, the Wyckoff position (0.25, 0.25, 0.25) is jointly occupied by Fe and Mn1, while Si and Mn2 occupy the (0.5,0.5,0.5) and (0,0,0) positions, respectively. By changing the structural model from $XA$ to L2$_1$, the refinement of the pattern gives the general formula to be  Mn$_{1.69}$Fe$_{1.31}$Si, which matches well with the target stoichiometry, as well as with the EDX results. This confirms that Mn$_{1.7}$Fe$_{1.3}$Si crystallizes in long-range L2$_1$ order. The refinement of the room temperature ND pattern with Fm$\bar{3}$m space group is shown in Fig\ref{fig: Fig.4}(a), and the crystallographic parameters obtained from the refinement are listed in Table-\ref{Table_1}. The inset of Fig.\ref{fig: Fig.4}(a) presents the crystal structure of Mn$_{1.7}$Fe$_{1.3}$Si after refinement.

\begin{table}[t]
  \begin{center}
    \caption{Crystallographic parameters of Mn$_{1.7}$Fe$_{1.3}$Si obtained from the Rietveld refinement of the neutron powder diffraction data at 300 K (space group Fm$\bar{3}$m, No. 225). The estimated lattice parameter is 5.6757(4) \AA.} 
    \label{Table_1}
    \setlength{\tabcolsep}{8pt}
    \renewcommand{\arraystretch}{1.5}
     \begin{tabular}{l c c c c r}\\
     \hline
        Atom & $x$ & $y$ & $z$ & Occ. & B$_{iso}$(\AA$^2$)  \\
        \hline
        Fe & 0.25 & 0.25 & 0.25 & 0.0270 & 0.487(0)\\
        Mn$1$ & 0.25 & 0.25 & 0.25 & 0.0146 & 0.487(0)\\
        Mn$2$ & 0.00 & 0.00 & 0.00 & 0.0211 & 0.629(25)\\
        Si & 0.50 & 0.50 & 0.50 & 0.0207 & 0.492(23)\\
    \hline
    \end{tabular}
  \end{center}
 \end{table}

Next, we consider the magnetic contribution to the NPD pattern. Fig.\ref{fig: Fig.3}(a) shows that the overall NPD pattern remains unchanged as temperature varies from 20 K to 300 K, confirming that there is no nuclear phase transition down to 20 K. Viewing from the high-temperature side, no additional Bragg reflections appear as the temperature reaches 70 K (i.e., $\leq$ $T_C$). The magnetic order in this temperature range can be weak FM/FiM, as there is no enhancement of the peak intensity below T$_C$, which may be due to the weak magnetic signal beyond the instrument's sensitivity. For further clarity, the temperature dependence of the integrated intensity of some selected Bragg peaks is shown in Fig.\ref{fig: Fig.3}(b), which reveals no intensity enhancement at low temperatures, preventing us from refining the magnetic structure at $T_C$. Upon lowering the temperature further, new satellite peaks appear at T $\leq$ 55 K, confirming the onset of AFM order. Notably, no AFM transition is detected between $T_{N1}$ and $T_{N2}$. This result counters the observation made in Ref.\cite{Mn1.7_2, Mn1.7_C}. It is worth noting that such a cusp-like feature in magnetization is often interpreted as the spin reorientation transition (SRT) in various Mn-based intermetallic alloys\cite{SRT-NiMnGa, SRT-NiMnGa-AM, SRT-Mn2RhSn, SRT-Ru2MnSb}. The SRT leads to a noncollinear magnetic order with uniaxial anisotropy and extreme sensitivity to the DC-field bias.

To account for the observed diffraction pattern below $T_{N1}$, we superimpose the nuclear structure in the space group Fm$\bar{3}$m with a magnetic structure where the magnetic moment is exclusive to the Mn sites. The magnetic propagation vector is of the form $k$ = [1/2,1/2,1/2], which is consistent with the previous report on Fe$_{3-x}$Mn$_x$Si\cite{Yoon_1977}. The refinement of the magnetic structure, shown in Fig.\ref{fig: Fig.4}(b), suggests a canted antiferromagnetic ordering, comprising of magnetic moments parallel to $<$111$>$ axis. These moments are ferromagnetically aligned within the respective $\{$111$\}$ planes while adjacent planes are coupled antiferromagnetically. The magnetic moment per Mn atom was determined to be 0.5413 $\pm$ 0.019 $\mu_B$. 
\begin{figure}[]
\centering
\includegraphics[width=\linewidth]{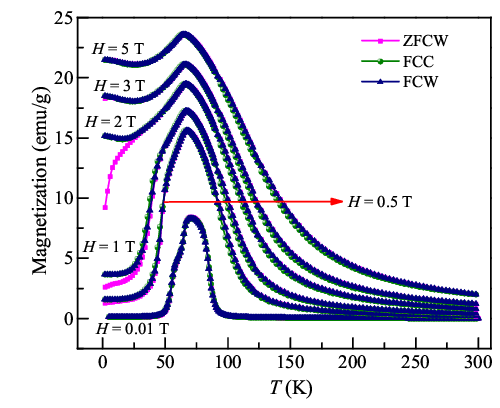}
 \caption{Variation of magnetization with the temperature ($M(T)$) measured at different applied magnetic fields under ZFC, FCC, and FCW protocols.}
    \label{fig: Fig.5}
\end{figure}

Fig.\ref{fig: Fig.5} shows the temperature-dependent magnetization, $M(T)$, recorded for different applied fields measured from 2 K to 300 K under ZFCW, FCC, and FCW protocols. A sharp rise in magnetization observed around 100 K is attributed to the magnetic phase transition from PM to FM/FiM state. Intriguingly, magnetization in this FM/FiM state decreases gradually after attaining a peak value at around 75 K. However, the corresponding NPD data shows only one AFM phase below 55 K. Therefore, the gradual decrease in magnetization below 75 K could be a result of a SRT from FM/FiM structure to canted AFM structure. This onset of SRT can be linked to the 
appearance of kink-like feature at $T$ $\sim$ 65 K in the $M(T)$ curve, as clearly marked in Fig.\ref{fig: Fig.2}. Additionally, below 55 K, the $M(T)$ curve becomes constant with a value close to zero, indicating the SRT occurs between 65 K and 55 K. 

\begin{figure}[]
 \includegraphics[width=\linewidth]{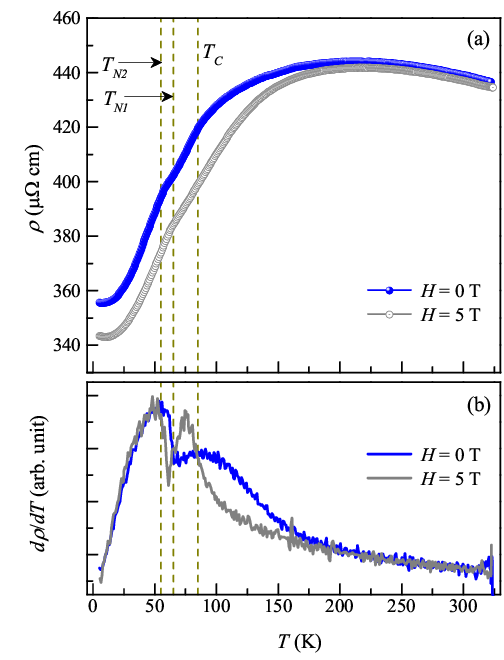}
 \caption{(a) Temperature-dependent resistivity, $\rho$(T) of Mn$_{1.7}$Fe$_{1.3}$Si and, (b) a plot derivative of $\rho$ with temperature under an applied magnetic field of 0 T and 5 T.}
    \label{fig: Fig.6}
    \end{figure}  

Further observations can be made from $M(T)$ curves measured at higher fields: (i) the bifurcation between the ZFCW and FCC curve first increases and then decreases with an increase in the applied field, as shown in the Fig.\ref{fig: Fig.5}; (ii) the kink around 65 K becomes less pronounced and eventually disappears at higher fields; and (iii) the step-like characteristics of $M(T)$ at around T$_{N1}$ at lower field gives way to regular peak-like characteristic at higher fields, which further becomes less pronounced and transition becomes more diffuse. The weakening of the feature associated with SRT and the diminishing of the bifurcation between ZFCW-FCC suggests that the SRT from the FM/FiM to the canted AFM structure is stable only in moderate fields and becomes weaker in higher fields. 

\begin{figure*}[]
\centering
\includegraphics[width=0.8\textwidth]{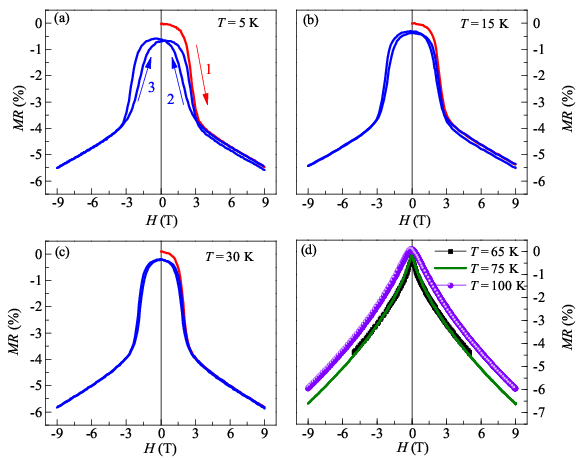}
 \caption{The field dependent transverse magnetoresistance (I $\perp$ B) of Mn$_{1.7}$Fe$_{1.3}$Si measured at fixed temperatures by varying the applied magnetic field of $\pm$ 9 T.}
    \label{fig: Fig.7}
\end{figure*}

Fig.\ref{fig: Fig.6}(a) presents the variation of electrical resistivity, $\rho$(T), with temperature measured under an applied magnetic field of 0 and 5 T. The $\rho$(T) curve at $H$ = 0 T shows three consecutive slope changes as temperature decreases. These anomalies are prominent in the derivative plots (Fig.\ref{fig: Fig.6}(b)), and the transition temperatures are in good agreement with those derived from magnetization measurements (at Fig.\ref{fig: Fig.2}). With decreasing temperature, $\rho(T)$ drops rapidly, followed by a gradual saturation to a constant value near 5 K, confirming the metallic character of the system.  Further, the $\rho$(T) measured with $H$ = 5 T begins to deviate from the 0 T curve around 150 K, indicating the presence of magnetoresistance over a wide temperature range. The effect of the magnetic field on magnetic transition temperatures can be observed in the derivative plots presented in Fig.\ref{fig: Fig.6}(b).

The corresponding field-dependent transverse (I $\perp$ B) magnetoresistance ($MR$) at various temperatures is shown in Fig.\ref{fig: Fig.7}(a)-(d). The $MR$ plots were obtained after warming the sample to the PM state before each measurement. At $T$ = 5 K, $MR$ remains nearly constant in the low field region ($\leq$ 1.5 T) and then decreases abruptly after $H_1$ $\sim$ 1.5 T. This sharp decrease in $MR$ persists until $H_2$ $\sim$ 3 T; thereafter, it decreases linearly with the field. Upon lowering the applied field (path-2), $MR$ retraces the curve up to $H_2$ but shows hysteresis with a further decrease in the field. During the field cycling from +9 T to -9 T along path-2 and -9 T and +9 T along path-3, the virgin curve (path-1) lies outside the envelope curve, forming a butterfly-like hysteresis loop. The $MR$ at $H$ = 0 after field excursion, denoted as $MR(0)$$^{mixed}$, is marginally lower than the initial or virgin $MR(0)$ value. As the temperature increases to 15 K and 30 K, the size of the butterfly-like loop becomes smaller compared to that observed at 5 K, and the $MR(0)$$^{mixed}$ value undergoes gradual reduction relative to the respective virgin $MR(0)$ value. With further increase in temperature, the hysteresis loop completely vanishes, and a quasi-linear behavior of the $MR$ curve is observed, as shown in Fig \ref{fig: Fig.7}(d). 

\begin{figure*}[]
\centering
\includegraphics[width=0.8\textwidth]{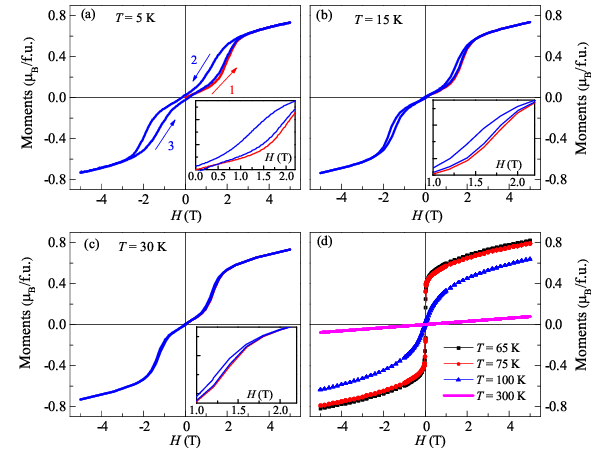}
 \caption{Isothermal field-dependent magnetization curves measured at fixed temperatures by varying the applied magnetic field of $\pm$ 5 T.}
    \label{fig: Fig.8}
\end{figure*}

This motivated us to investigate the isothermal field-dependent magnetization, $M(H)$. The $M(H)$ data were measured at selected temperatures between 5 K and 300 K over a field range of $\pm$5 T, as shown in Fig.\ref{fig: Fig.8} (a-d). In the low-field region, $M(H)$ shows linear variation with the field, akin to AFM ordering of the material. However, an increase in $M$ at a specific field value, $H_1$, indicates a field-induced metamagnetic transition, occurring over a broad field range from $H_1$ ($\sim$ 1.5 T) to $H_2$ ($\sim$ 3 T). As the field increases further from 3 to 5 T, the $M(H)$ curve shows a curvature with a non-saturating tendency. A hysteresis is observed when the field is reduced from 5 T to 0 T. The hysteresis loop, however, collapses before the field reaches zero, exhibiting a high coercive field of 0.17 T. This $M(H)$ behavior is irreversible following the first-order magnetic transition. The $M(H)$ recorded at 15 K and 30 K also show similar behavior, though the metamagnetic transitions and hysteresis loop occur at lower field values. The virgin curve lying outside the envelope loop and the irreversible hysteresis suggest an incomplete metamagnetic transition for T $\leq$ 30 K. However, the $M(H)$ recorded at $T$ $\geq$ 65 K becomes fully reversible in all quadrants, with the virgin curve coinciding with the corresponding part of the hysteresis loop. Additionally, the feature of metamagnetic transition completely vanishes at 65 K, and a soft FM/FiM behavior is observed. A paramagnetic $M(H)$ is observed at 300 K. Returning to the description of the $M(H)$ observed at $T$ $\leq$ 30 K, the smooth curvature of the high field, non-saturating $M(H)$ with high coercivity confirms the FM/FiM ordering of the high-field phase. This high-field phase is partially retained even when the field is reduced to zero, suggesting a coexistence of AFM--FM/FiM phase in the material. The influence of these magnetic transitions is echoed in the $MR$ measurements discussed earlier. 

\section{Discussion}
\subsection{\textit{\textbf{Spin-reorientation transition and canted antiferromagnetic ordering}}}
\begin{figure}[]
 \includegraphics[width=\linewidth]{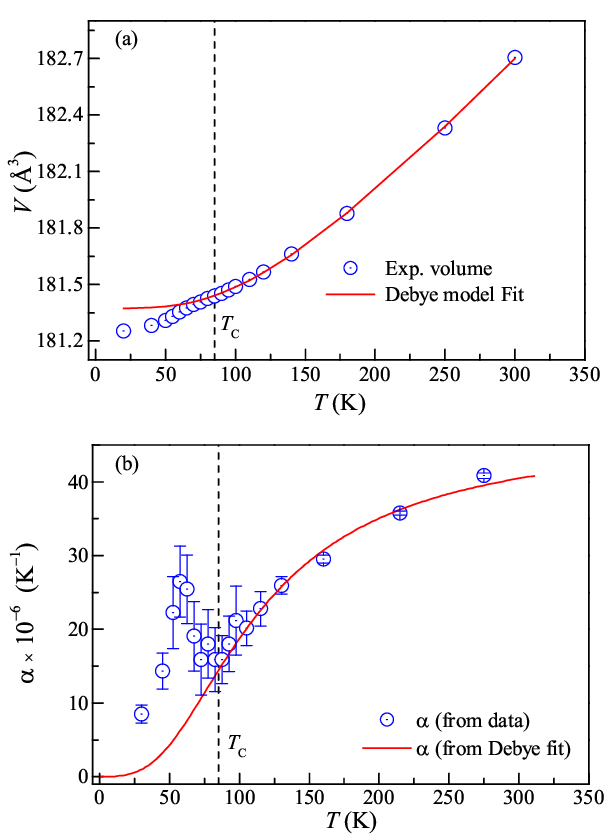}
 \caption{(a) The variation of unit cell volume with temperature, where the solid line shows the fit obtained using the Debye-Grüneisen model. (b) Shows the calculated volumetric thermal expansion coefficient, $\alpha$ (open symbols), alongside its theoretical variation (red curve) expected from the Debye fit .}
    \label{fig: Fig.9}
    \end{figure}

Spin-reorientation transition (SRT) has been studied extensively in perovskites, rare-earth oxides, and intermetallic compounds\cite{SRT-1, SRT-2, SRT-3, SRT-4, SRT-NiMnGa, SRT-NiMnGa-AM}. A prerequisite for SRT is the presence of magnetostructural coupling. Fig.\ref{fig: Fig.9}(a) shows the temperature dependence of the lattice volume, $V(T)$, derived from the Rietveld refinement of the NPD data. Although the cubic symmetry is preserved down to 20 K, the unit cell volume shows a clear slope change near $T_C$. The observed changes in unit cell volume around the magnetic transitions, without any change in crystallographic symmetry, suggest an isostructural distortion driven by magnetoelastic coupling. However, no prior studies have reported this effect in Mn$_{1.7}$Fe$_{1.3}$Si, making its investigation crucial for uncovering the mechanisms underlying the SRT in this compound. The magnetoelastic coupling can be validated by modeling the thermal expansion of unit cell volume using the Debye-Gr$\ddot{\mathrm{u}}$neisen model:
\begin{equation}
    V(T) = V_0 + \frac{9 \gamma k_BT}{B} \left(\frac{T}{\Theta_D}\right)^3\int_{0}^{\Theta_D/T}\frac{x^3}{e^x - 1}\,dx
\end{equation}

where $V_0$, $\gamma$, $k_B$, B, and $\Theta_D$ are unit cell volume at 0 K, Grüneisen parameter, Boltzmann constant, bulk modulus, and Debye temperature, respectively. The extracted fitting parameters are $V_0$ = (181.3739$\pm$0.0006)\AA$^3$ and $\Theta_D$ = 460.66$\pm$4.45 K, which is reassuringly close to the value obtained from specific heat measurement\cite{Mn1.7_C}. It is evident from Fig.\ref{fig: Fig.9}(a) that above $T_C$, the experimental unit cell volume is well described by the Debye-Gr$\ddot{\mathrm{u}}$neisen model. However, large deviations occur in the vicinity of magnetic transitions, indicating a need for alteration in the underlying structural model. These observations confirm the genesis of the magnetic transitions in Mn$_{1.7}$Fe$_{1.3}$Si to be of magnetoelastic coupling-induced structural distortion\cite{MEC1,MEC2,MEC3}. To further support this conclusion, we calculated the volumetric thermal expansion coefficient ($\alpha$ = $1/V(dVdT)$). Fig.\ref{fig: Fig.9}(b) shows the temperature variation of $\alpha$ alongside the expected trend from the Debye-Gr$\ddot{\mathrm{u}}$neisen model. The observed deviation in $\alpha$ from its expected theoretical behavior below $T_C$ highlights the presence of strong magneto-elastic coupling in the system. 

Previous NPD studies on Fe$_{3-x}$Mn$_x$Si\cite{Yoon_1977} have observed new satellite peaks at 4.2 K but failed to capture any change in the NPD profile between $T_{N1}$ and $T_{N2}$, probably due to large temperature intervals used in those measurements. In contrast, the present study utilizes high-resolution NPD patterns recorded at an interval of 5 K below $T_C$, enabling us to capture anomalies in lattice parameters across the $T_C$ and SRT. Our results show that the SRT in Mn$_{1.7}$Fe$_{1.3}$Si is correlated with marked structural distortions around magnetic transition temperature. 

\subsection{\textit{\textbf{Metamagnetic transition and its effect on the magnetic and transport properties}}}
\begin{figure}[]
 \includegraphics[width=\linewidth]{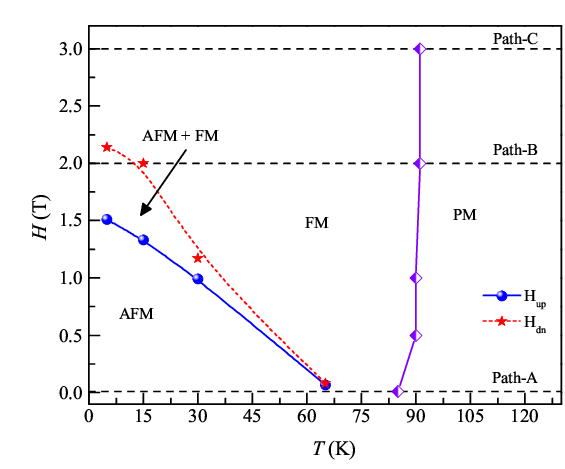}
 \caption{The $H-T$ phase diagram of Mn$_{1.7}$Fe$_{1.3}$Si.}
    \label{fig: Fig.10}
    \end{figure}

Based on the experimental observations made in this study, an $H-T$ phase diagram of Mn$_{1.7}$Fe$_{1.3}$Si has been constructed, as shown in Fig.\ref{fig: Fig.10}. The critical magnetic field values for the field-induced metamagnetic transition, while increasing ($H_{up}$) and decreasing ($H_{dn}$), are plotted against temperature. FM/FiM and AFM phases coexist between $H_{up}$ and $H_{dn}$.

Tracing the anomalous magnetic behavior of Mn$_{1.7}$Fe$_{1.3}$Si through this phase diagram, Path-$A$, $B$, and $C$ indicate the $M(T)$ measurements at various constant fields. When the sample is cooled in zero field, it transforms entirely into the AFM phase. Following Path-$A$ (at a small applied magnetic field) under ZFCW, FCC, and FCW protocols results in almost no bifurcation, as it lies outside the phase-coexistence region. A similar scenario is observed for Path-$C$ (for $H$ $\geq$ 2.5 T). However, the low-temperature portion of Path-$B$ lies within the phase-coexistence region. Cooling and warming the sample in the 0 and 2 T leads to bifurcation in the ZFCW and FCC/FCW curves. The absence of bifurcation in the ZFCW and FCC curve in the low-$T$ low-$H$ and low-$T$ high-$H$ regions confirms the long-range single phase of Mn$_{1.7}$Fe$_{1.3}$Si. This observation also supports our isothermal $M(H)$ data, where the low-$H$ AFM region transforms into the FM/FiM state as the magnetic field increases. As discussed earlier, the initial cooling of the sample in a 2 T field results in a finite fraction of the high-temperature FM/FiM phase to exist even at $T$ = 5 K, highlighting the presence of hindered kinetics during the phase transformation. During isothermal field reduction from 5 T to 0 T, the kinetics of the FM/FiM to AFM transition become impeded, indicating metastable behavior in Mn$_{1.7}$Fe$_{1.3}$Si at or below 5 K after field cycling. To the best of our knowledge, unlike many intermetallic compounds \cite{Nd7Rh3, Mn2Sb}, non-cubic Heusler alloys\cite{Mn2PtGa, Ni-Mn-In, Ni-Mn-Sn}, and manganites\cite{NdMnO3}, the phenomenon of field-induced magnetic phase transitions accompanied by hindered kinetics and metastability has not been reported in cubic Heusler compounds. Such behavior, however, is observed in germanides due to their coupled crystallographic-magnetic transition\cite{Gd5Ge4-1, Gd5Ge4-2, Gd5Ge4-3, Gd5Ge4-4}.

Next, we address the nature of $MR$ data to investigate the effect of metastability, kinetic hindrance, and phase co-existence on transport properties. Referring to Fig.\ref{fig: Fig.7}, the $MR$ plots measured at high temperatures (65 K, 75 K, and 100 K) exhibit a negative magnetoresistance, which decreases with the field starting from $H$ = 0. At lower temperatures, the virgin curves of the $MR$ plots (path-1) showed a field-induced AFM to FM/FiM transition. The decrease in $MR$ begins at $H_1$ due to the onset of the FM/FiM phase, as the AFM phase gradually converts into the FM/FiM phase. This process continues until the field reaches $H_2$, beyond which $MR$ exhibits the typical negative behavior associated with the fully transformed FM/FiM state across the entire sample at high fields. During the field decreasing cycle (path-2), the entire sample remains in the phase-coexisted form with the lower $MR(0)$ compared to the virgin $MR(0)$. This is an expected behavior across the first-order transition, indicating metastability when the field is cycled across the transition field. The anomalous open hysteresis loop in the $MR$ is apparent as the virgin curve lies outside the butterfly-like loop. The difference between the $MR(0)^{mixed}$ and the corresponding virgin $MR(0)$ gradually decreases with increasing temperature, reflecting the reduction in the size of the butterfly loop compared to that at 5 K. These observations are consistent with our $M(H)$ data. 

\section{Conclusion}
In conclusion, a comprehensive study of structural, magnetic, and transport properties of cubic Mn$_{1.7}$Fe$_{1.3}$Si has been conducted. The room temperature powder XRD pattern confirms a long-range cubic L2$_1$ order, while temperature-dependent NPD further supports the cubic Heusler structure without any change in crystal symmetry down to 20 K. The magnetic structure refinement at $T$ = 20 K confirms a canted antiferromagnetic ordering. The variation in magnetization below 70 K is attributed to a spin reorientation transition, likely driven by a gradual change in magnetocrystalline anisotropy, as evidenced by the unusual behavior of the unit cell volume around magnetic transition temperatures. The $\rho$(T) measurement confirms the low-temperature metallic behavior of Mn$_{1.7}$Fe$_{1.3}$Si. Furthermore, we have established the presence of field-induced metamagnetic transition at low temperatures. Several features characterizing the first-order transition, such as irreversibility, metastability, and phase-coexistence, are seen distinctly in the magnetic and magnetotransport properties. While the characteristics of the compound are intriguing, the question remains whether the first-order magnetic transition is driven by a simultaneous structural transition as seen in other intermetallic compounds\cite{Gd5Ge4-3} and shape memory Heusler alloys\cite{Ni-Mn-Sn, Ni-Mn-In}. Therefore, further neutron and X-ray diffraction measurements as a function of the magnetic field will be highly valuable in addressing this question.

\section{Acknowledgment}
K.M. acknowledges the DST-INSPIRE fellowship (File: DST/INSPIRE/03/2021/001384/IF190777), from the Department of Science and Technology, New Delhi, for providing the research fellowship. We also thank Prof. Pratap Raychaudhuri and Mr. Ganesh Jangam for their assistance with magnetic measurement at the Tata Institute of Fundamental Research, Mumbai. The authors acknowledge the Science and Technology Facility Council (STFC UK) and ISIS facility for providing neutron beam time (RB2220003) on the HRPD beamline.

\bibliography{mybib}
\end{document}